\documentclass[12pt]{article}
\usepackage{amsmath}

\newcommand{\ben}{\begin{enumerate}}
\newcommand{\een}{\end{enumerate}}
\newcommand{\be}{\begin{equation}}
\newcommand{\ee}{\end{equation}}
\newcommand{\bse}{\begin{subequation}}
\newcommand{\ese}{\end{subequation}}
\newcommand{\bea}{\begin{eqnarray}}
\newcommand{\eea}{\end{eqnarray}}
\newcommand{\bc}{\begin{center}}
\newcommand{\ec}{\end{center}}

\begin{document}

\begin{center}

\textsf{\LARGE Applications of Classical and Quantum Algebras to
Molecular Thermodynamics}

 \vspace{7mm}

\textbf{Maia Angelova}

\vspace{5mm}

\textit{School of Computing and Mathematics}\\ \textit{University
of Northumbria, Newcastle upon Tyne, UK GB-NE1 8ST}\\
\end{center}

\vspace{.3 cm}

\begin{abstract}
Lie-algebraic and quantum-algebraic techniques are used in the
analysis of thermodynamic properties of molecules and solids. The
local anharmonic effects are described by a Morse-like potential
associated with the $su(2)$ algebra. A vibrational
high-temperature partition function and the related thermodynamic
potentials are derived in terms of the parameters of the model.
Quantum analogues of anharmonic bosons, $q$-bosons, are introduced
and used to describe anharmonic properties of molecules and
solids. It is shown that the quantum deformation parameter is
related to the fixed number of anharmonic bosons and the shape of
the anharmonic potential. A new algebraic realization of the
$q$-bosons, for the case of $q$ being a root of unity is given.
This realization represents the symmetry of a linear lattice with
periodic boundary conditions.
\end{abstract}

%
%
%
%
%
%
%
%
\section{Introduction}
The algebraic approach has been developed as an alternative to
{\it ab initio} and Dunham-like approaches to describe the
molecular vibrational degrees of freedom \cite
{{frank:algebraic},{iach:algebraic}}. The algebraic approach
provides the energy as an analytic function of the quantum
numbers. The Hamiltonian is written as an algebraic operator using
the appropriate Lie algebras. The technical advantage of the
algebraic approach is the comparative ease of the algebraic
operations. Equally important is the result obtained by comparison
with the experiment, that there are generic forms of algebraic
Hamiltonians and that entire class of molecules can be described
by common Hamiltonian where only the parameters are different for
different molecules. In its initial stage of development
\cite{iach:chem}-\cite{lemus:chem}, the algebraic approach has
sought to show why and how it provides a framework for the
understanding of large-amplitude anharmonic motion. The
anharmonicities are introduced by means of dynamical groups that
correspond to anharmonic potentials and which constrain the total
number of levels to a finite value. The current algebraic models
\cite{frank:annals}-\cite{lemus:99} combine Lie algebraic
techniques, describing the interatomic interactions, with discrete
symmetry techniques associated with the local symmetry of the
molecules. In the anharmonic oscillator symmetry model
\cite{frank:annals}, the local internal coordinates are given in
terms of $u(2)$ algebras. The $u(2)$ interactions correspond to
anharmonic coupling of anharmonic oscillators which approximate
the interactions between the Morse oscillators.

The algebraic anharmonic model has been developed to analyze molecular
vibrational spectra \ \cite{frank:algebraic}-\cite{lemus:99}. It provides a
systematic procedure for studying vibrational excitations in a simple form
by describing the stretching and bending modes in a unified scheme based on $%
u(2)$ algebras which incorporate the anharmonicity at the local
level.

The success of the algebraic models in the analysis of molecular
vibrational spectra has led to the development of similar models
for molecular thermodynamics. Kusnetzov \cite{kusnetzov:chem} has
used an algebraic approach, based on approximation of the
classical density of states, to study thermodynamic properties of
polyatomic molecules. Angelova and Frank
\cite{{angelova:trjphys},{angelova:CMP},{angelova:Kra}} have
applied the algebraic model
\cite{{frank:algebraic},{frank:annals}} to the vibrational
high-temperature thermodynamics of diatomic molecules and derived
the vibrational partition function and the important thermodynamic
functions, such as mean energy and specific heat, in terms of the
parameters of the model.

The Morse-like potential, which represents the anharmonicities at
the local level, leads to a deformation of the harmonic oscillator
algebra. Angelova, Dobrev and Frank \cite{angelova:JPA} have
derived this deformation using quantum analogue of the anharmonic
oscillator. We have described the anharmonic vibrations as
anharmonic $q$-bosons using first-order of the expansion of a
quantum deformation and found relations between the parameters of
the algebraic model and the quantum deformation parameter.

The aim of this paper is to review the applications of the
classical and quantum algebras to molecular thermodynamics. In
Section 2, the framework of the algebraic model is given. The
vibrational  partition function and the related thermodynamic
functions, such as mean energy, specific heat and the mean number
of the anharmonic bosons, are discussed in Section 3. The
anharmonic $q$-bosons are discussed in Section 4. A $q$-bosonic
deformation of first order is considered and it is shown that the
corresponding  quantum deformation parameter is related to the
shape of the anharmonic potential well and the fixed number of
anharmonic bosons. The $q$-bosons at roots of unity, which give
rise to a finite-dimensional periodic structure are discussed and
their applications to a linear lattice with periodic boundary
conditions are given.

\section{Algebraic Model}
The algebraic model \cite{frank:algebraic} exploits the
isomorphism of $su(2) $ algebra and the one-dimensional Morse
oscillator. The one-dimensional Morse Hamiltonian can be written
in terms of the generators of $su(2)$,
\begin{equation}
\mathcal{H}_{M}=\frac{A}{4}\left( \hat{\mathcal{N}}^{2}-4\hat{J}_{Z}^{2}\right) =\frac{%
A}{2}(\hat{J}_{+}\hat{J}_{-}+\hat{J}_{-}\hat{J}_{+}-\hat{\mathcal{N}})
\end{equation}
where $A$ is a constant dependent on the parameters of the Morse
potential. The eigenstates, $|[\mathcal{N}\!],v\rangle $,
\begin{equation}
|\![\mathcal{N}],v\rangle = \sqrt{\frac{\left( \mathcal{N}-v\right) !}{\mathcal{%
N}!v!}}\left( J_{-}\right) ^{v}|[\mathcal{N}],0\rangle
\end{equation}
correspond to the $\ u(2)\supset su(2)$ symmetry-adapted
basis,$\;$where $\mathcal{N}\;$is the total number of bosons fixed
by
the potential shape, and $v$ is the number of quanta in the oscillator, $%
v=1,2,\ldots ,\left[ \frac{\mathcal{N}}{2}\right] $. These wave
functions can be further symmetry-adapted to the local symetry of
the molecules in a way described for example in
\cite{frank:annals}.

The value of $\mathcal{N}$ is dependent on the depth $D$ and the
width $d$ of the Morse potential well \cite{frank:annals},
\begin{equation}
\mathcal{N}+1=\left( \frac{8\mu Dd^{2}}{\hbar ^{2}}\right)
^{\frac{1}{2}}. \label{ND}
\end{equation}
where $\mu$ is the mass of the oscillator. The parameters $A$ and
$\mathcal{N}$ are related to the usual harmonic and anharmonic
constants $\omega_{e}$ and $x_{e}\omega_{e}$ used in spectroscopy
\cite{{iach:chem},{levine:chem},{herzberg:spectra}},
\begin{align}\label{wx}
\omega_{e}  &  =A(\mathcal{N}+1)=\hbar\left(  \frac{2D}{\mu d^{2}}\right)  ^{\frac{1}%
{2}}\\ x_{e}\omega_{e}  &  =A=\frac{\hbar^{2}}{2d^{2}D}\nonumber
\end{align}

The anharmonic effects are described by anharmonic boson operators
\cite {frank:algebraic},
\begin{equation}
\hat{b}=\frac{\hat{J}_{+}}{\sqrt{\mathcal{N}}},\;\;\;\hat{b}^{\dagger }=\frac{\hat{J}%
_{-}}{\sqrt{\mathcal{N}}}\;,\;\;\;\hat{v}=\frac{\hat{\mathcal{N}}}{2}-\hat{J}_{z}
\label{anhb}
\end{equation}
where $\hat{v}\;\;$is the Morse phonon operator with an eigenvalue
$v$. The operators satisfy the commutation relations,
\begin{equation}
\left[ \hat{b},\hat{v}\right] =\hat{b},\;\;\;\;\;\;\;\left[ \hat{b}%
^{^{\dagger }},\hat{v}\right] =-\hat{b}^{^{\dagger }},\;\;\;\;\left[ \hat{b},%
\hat{b}^{^{\dagger }}\right] =1-\frac{2\hat{v}}{\mathcal{N}}
\label{anh}
\end{equation}
The harmonic limit is obtained when $\;\;\mathcal{N}\rightarrow
\infty $, \ in which case \ \ $\left[ \hat{b},\hat{b}^{^{\dagger
}}\right] \!\!\rightarrow \!\!1$ \ giving the usual boson
commutation relations.

The Morse Hamiltonian can be written in terms of \ the operators
$\hat{b}$ and $\hat{b}^{^{\dagger}}$,
\begin{equation}
H_{M}\sim\frac{1}{2}\left( \hat{b}\hat{b}^{^{\dagger}}+\hat{b}^{^{\dagger}}%
\hat{b}\right)
\end{equation}
which corresponds to vibrational energies
\begin{equation}
\varepsilon_{v}=\hbar\omega_{0}\left(
v+\frac{1}{2}-\frac{v^{2}}{\mathcal{N}}\right)
,\;v=1,2,\ldots,\left[ \frac{\mathcal{N}}{2}\right]   \label{Ev}
\end{equation}
where $\omega_{0}$ is the harmonic oscillator frequency. Thus, the
spectrum of the Morse potential leads to a deformation of the
harmonic oscillator algebra. A more detailed relationship between
the Morse coordinates and momenta and the $su(2)$ generators can
be derived through a comparison of their matrix elements
\cite{lemus:99}. Note that for an infinite potential depth,
$\mathcal{N}\!\rightarrow\!\infty$, the Morse potential cannot be
distinguished from the harmonic potential.

\section{Thermodynamic Vibrational Functions}
\subsection{Vibrational Partition Function}

In the anharmonic algebraic approach, the vibrational partition
function of a diatomic anharmonic molecule is
\begin{equation}
Z_{\mathcal{N}}=\sum_{v=0}^{[\mathcal{N}/2]}e^{-\beta\varepsilon_{v}}
\end{equation}
where $\beta=1/k_{B}T$, the vibrational energies $\varepsilon_{v}$
are given by equation (\ref{Ev}) and $\mathcal{N}$ is the fixed
total number of anharmonic
bosons discussed in the previous section.  Introducing new parameters,$\;\alpha=\frac{\beta\hbar\omega_{0}}{2}%
,\;\mathcal{N}_{0}=\left[ \frac{\mathcal{N}}{2}\right] $ and
$\;l=\mathcal{N}_{0}-v$, the exact value of vibrational partition
function can be written as,
\begin{equation}
Z_{\mathcal{N}}=e^{-\alpha}\sum_{l=0}^{\mathcal{N}_{0}}e^{-\frac{\alpha}{\mathcal{N}_{0}}\left(
\mathcal{N}_{0}^{2}-l^{2}\right) }.   \label{exactpf}
\end{equation}
At high temperatures $T$, for $\mathcal{N}_{0}$ large and $\alpha$
small, the sum can be replaced by the integral,
\begin{equation}
Z_{\mathcal{N}}=\sqrt{\frac{\mathcal{N}_{0}}{\alpha}}e^{-\alpha\left(
\mathcal{N}_{0}+1\right) }\int\limits_{0}^{\sqrt{\alpha
\mathcal{N}_{0}}}e^{s^{2}}ds
\end{equation}
where $s=\sqrt{\frac{\alpha}{\mathcal{N}_{0}}}l$. This integral
can be evaluated exactly in terms of the error function,
${\hbox{\rm erf}\,}i\!\left( \sqrt{{\alpha}\mathcal{N}_{0}}\right)
$ (as defined in \cite {abram:handbook}),
\begin{equation}
Z_{\mathcal{N}}=\frac{1}{2}\sqrt{\frac{\mathcal{N}_{0}{\pi}}{\alpha}}e^{-\alpha\left(
\mathcal{N}_{0}+1\right) }{\hbox{\rm erf}\,}i\!\left( \sqrt{\alpha
\mathcal{N}_{0}}\right) . \label{pf}
\end{equation}
Equation (\ref{pf}) represents the high-temperature value of the vibrational
partition function in the Morse-like spectrum. The partition function is
expressed in terms of the parameters of the algebraic model $\mathcal{N}_{0}$ and $%
\alpha$. When $\mathcal{N}_{0}\rightarrow\infty $, the harmonic
limit of the model is obtained,
\begin{equation}
Z_{\infty}\sim\frac{\mathcal{N}_{0}e^{-\alpha}}{2\alpha \mathcal{N}_{0}-1}\sim\frac{e^{-\alpha}}{%
2\alpha}=\frac{T}{\Theta}e^{-\frac{\Theta}{2T}}   \label{Zhl}
\end{equation}
which coincides with the harmonic vibrational partition function
of a diatomic molecule at high temperatures. Here, $%
\Theta=\hbar\omega_{0}/k_{B}$, is the usual characteristic
vibrational temperature of the molecule and the model parameter
$\alpha={\Theta}/2T$.  The expression for the partition function
(\ref{pf}) can be further generalised for polyatomic molecules.

\begin{figure}[t]
\input{epsf} \centerline{\epsffile{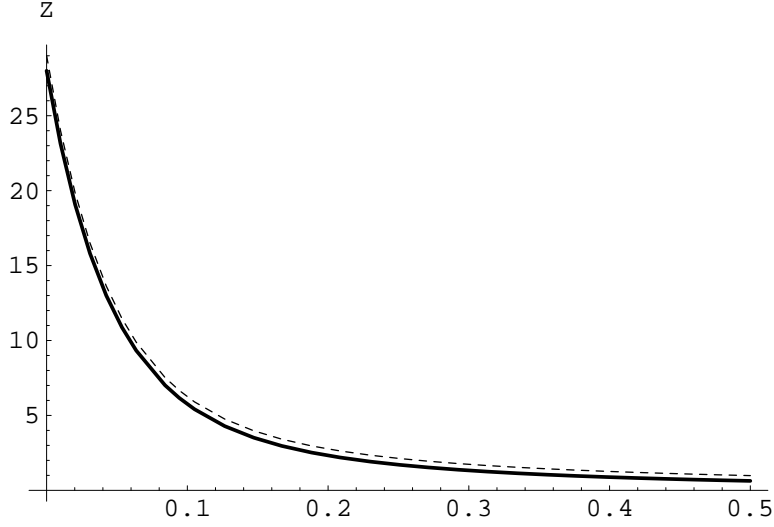}}
\caption{ Vibrational partition function $Z_{56}$ as a function of $\protect%
\alpha$. }
\label{fig:vibra}
\end{figure}

Using the values of the harmonic and anharmonic constants
\cite{herzberg:spectra} for the zero lines of the diatomic
molecule $^{1}$H$^{35}$Cl  and the equations (\ref{ND}) and
(\ref{wx}), we obtain the total number of anharmonic bosons, fixed
by the shape of the Morse
potential, $\mathcal{N}=56$, and the total number of quanta in the oscillator, $%
\mathcal{N}_{0}=28$. The characteristic vibrational temperature of
the molecule is $\Theta=4300$K.

Substituting the value of $\mathcal{N}_{0}=28$ in equation
(\ref{pf}) we can calculate the partition function, $Z_{56}$, for
the molecule $^{1}$H$^{35}$Cl as a function of the parameter
$\alpha$. The  anharmonic effects  are essential at the high
temperatures  $T\geq\Theta$, {\it i.e.} $\alpha\leq0.5$. The graph
on Figure \ref
{fig:vibra} represents the partition function $Z_{56}$ given by equation (%
\ref{pf}) for the values of the parameter $\alpha$ between 0 and
0.5 (solid line). The exact partition function from equation
(\ref{exactpf}) is given for comparison (dashed line). It is
clear, that the integral approximation is in a very good agreement
with the exact representation and does not change the value and
appearance of the partition function.

\subsection{Mean Vibrational Energy}

The mean vibrational energy is given by
\begin{equation}
U_{\mathcal{N}}=-\frac{\partial}{\partial\beta}\hbox{\rm
ln}Z_{\mathcal{N}}=-\frac{\hbar\omega
_{0}}{2Z_{\mathcal{N}}}\frac{\partial
Z_{\mathcal{N}\;}}{\partial\alpha}.
\end{equation}
Taking into account that
\begin{equation}
\frac{\partial
Z_{\mathcal{N}\;}}{\partial\alpha}=-\frac{Z_{\mathcal{N}}}{2\alpha}-\left(
\mathcal{N}_{0}+1\right)
Z_{\mathcal{N}}+\frac{\mathcal{N}_{0}e^{-\alpha}}{2\alpha}
\label{dZ}
\end{equation}
we obtain the following expression for the mean vibrational energy
in terms of the partition function $Z_{\mathcal{N}}$,

\begin{equation}
U_{\mathcal{N}}=\frac{\hbar\omega_{0}}{2}\left(
1+\mathcal{N}_{0}+\frac{1}{2\alpha}-\frac
{\mathcal{N}_{0}e^{-\alpha}}{2\alpha Z_{\mathcal{N}}}\right) .
\label{UN}
\end{equation}
Substituting $Z_{\mathcal{N}}$ with equation (\ref{pf}) gives the
following expression for the mean energy, $U_{\mathcal{N}}$, in
terms of the parameter ${\alpha}$,
\begin{equation}
U_{\mathcal{N}}=\frac{\hbar\omega_{0}}{2}\left( 1+\mathcal{N}_{0}+\frac{1}{2\alpha}- \sqrt {%
\frac{\mathcal{N}_{0}}{\alpha\pi}} \frac{e^{\alpha
\mathcal{N}_{0}}}{\hbox{\rm erf}\,i\left( \sqrt{\alpha
\mathcal{N}_{0}}\right) }\right) .   \label{U}
\end{equation}
The harmonic limit is obtained from equation (\ref{UN}), when $\mathcal{N}_{0} %
\rightarrow\infty$ and $Z_{\mathcal{N}}$ is given by (\ref{Zhl}),
\begin{equation}
U_{\infty}\sim\frac{\hbar\omega_{0}}{2}(1+\frac{1}{\alpha})\sim\frac
{\hbar\omega_{0}}{2}+k_{B}T.
\end{equation}
This is the classical mean energy of a diatomic molecule at very high
temperatures.

\subsection{Specific Heat}

The vibrational part of the specific heat is,
\begin{equation}
C_{\mathcal{N}}=\frac{\partial U_{\mathcal{N}}}{\partial T}=-\frac{\hbar\omega_{0}}{2k_{B}T^{2}}%
\frac{\partial U_{\mathcal{N}}}{\partial\alpha}.
\end{equation}
Substituting $U_{\mathcal{N}}$ with equation (\ref{UN}) and using
equation (\ref{dZ}), we obtain
\begin{equation}
C_{\mathcal{N}}=\frac{k_{B}}{2}+\frac{k_{B}\mathcal{N}_{0}e^{-\alpha}}{2Z_{\mathcal{N}}}\left(
\alpha
\mathcal{N}_{0}-\frac{1}{2}-\frac{\mathcal{N}_{0}e^{-\alpha}}{2Z_{\mathcal{N}}}\right)
\label{CN}
\end{equation}
The equation (\ref{CN})  represents the vibrational specific heat
in the algebraic model in terms of the partition function
$Z_{\mathcal{N}}$. Substituting $Z_{\mathcal{N}}$\ with (\ref{pf})
in the equation (\ref{CN}), we obtain the dependence of the
specific heat $C_{\mathcal{N}}$ on the parameters of the model
${\alpha}$ and $\mathcal{N}_{0}$,
\begin{equation}
C_{\mathcal{N}}=\frac{k_{B}}{2}+k_{B}\sqrt{\frac{\alpha
\mathcal{N}_{0}}{\pi}}\frac{e^{\alpha \mathcal{N}_{0}}}{{\hbox{\rm
erf}\,}i\left( \sqrt{\alpha \mathcal{N}_{0}}\right) }\left( \alpha
\mathcal{N}_{0}-\frac{1}{2}-\sqrt{\frac{\alpha \mathcal{N}_{0}}{\pi}}\frac{e^{\alpha \mathcal{N}_{0}}}{%
\hbox{\rm erf}\,i\left( \sqrt{\alpha \mathcal{N}_{0}}\right)
}\right) \label{C}
\end{equation}

When $\mathcal{N}_{0}~$ $\rightarrow\infty$, the harmonic limit of
the model gives the vibrational specific heat of a diatomic
molecule at very high temperatures,
\begin{equation}
C_{\infty}\sim k_{B}. \label{CHL}
\end{equation}

\subsection{Mean Number of Anharmonic Bosons}

\begin{figure}[t]
\input{epsf} \centerline{\epsffile{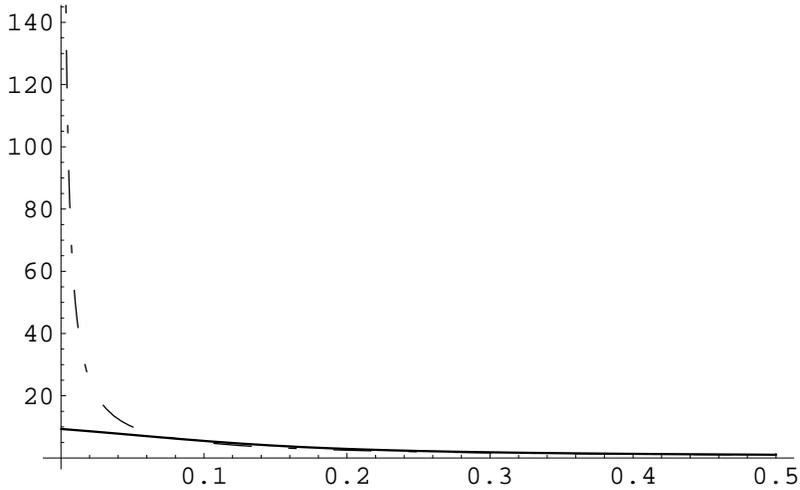}}
\caption{Mean Number of Anharmonic Bosons $\protect\nu _{56}$ as a function
of $\protect\alpha $ }
\label{fig:bosons}
\end{figure}

The mean vibrational energy in the anharmonic model can be written
in terms of mean number\ $\langle \nu _{\mathcal{N}}\rangle $ of
anharmonic quanta, each with energy $\hbar \omega _{0}$,

\begin{equation}
U_{\mathcal{N}}=\hbar\omega_{0}\left(
\langle\nu_{\mathcal{N}}\rangle+\frac{1}{2}\right) \label{Uv}
\end{equation}
Substituting $U_{\mathcal{N}}$ by equation (\ref{UN}), we obtain
$\langle\nu_{\mathcal{N}}\rangle $ in terms of the partition
function $Z_{\mathcal{N}},$
\begin{equation}
\langle\nu_{\mathcal{N}}\rangle=\frac{\mathcal{N}_{0}}{2}+\frac{1}{4\alpha}-\frac{%
\mathcal{N}_{0}e^{-\alpha }}{4\alpha Z_{\mathcal{N}}}. \label{vN}
\end{equation}
Using the expression (\ref{pf}) in equation (\ref{vN}), we obtain the
high-temperature value,
\begin{equation}
{\label{v}\langle\nu_{\mathcal{N}}\rangle=\frac{\mathcal{N}_{0}}{2}+\frac{1}{4\alpha}-\sqrt {%
\frac{\mathcal{N}_{0}}{4{\pi}{\alpha}}}\frac{e^{\alpha
\mathcal{N}_{0}}}{\hbox{\rm erf}\,i\left( \sqrt{\alpha
\mathcal{N}_{0}}.\right) }}
\end{equation}
The harmonic limit is obtained from equation (\ref{vN}) when $%
\mathcal{N}_{0}\rightarrow\infty$ and $Z_{\mathcal{N}}$ is given
by the expression (\ref{Zhl}),
\begin{equation}
\langle\nu_{\infty}\rangle\sim\frac{k_{B}T}{\hbar\omega_{0}}.
\end{equation}
The graph of the function $\langle\nu_{56}\rangle$ for the molecule $^{1}$H$%
^{35}$Cl is given on Figure \ref{fig:bosons} (solid line). The
harmonic limit $\langle\nu_{\infty}\rangle$ is given for
comparison (dashed line).

\section{Applications of q-Bosons}

\subsection{Anharmonic $q$-Bosons}

We have shown in \cite{angelova:JPA} that the anharmonic bosons $%
b,b^{\dagger }$ \ from equations  (\ref{anhb}) can be obtained as
an approximation of the $q$-bosons \cite
{arik-coon:jmp,biedenharn:jphysa,macfarlane:jphysa}. The
$q$-bosons, enter the Heisen\-berg-Weyl $q$-algebra $HW_{q}$ \ by
the following commutation relations:
\begin{equation}
\lbrack a,a^{\dagger }]=q^{\hat{n}}\ ,\quad \lbrack \hat{n},a]=-a\ ,\quad
\lbrack \hat{n},a^{\dagger }]=a^{\dagger }  \label{qanh}
\end{equation}
where the deformation parameter $q$ is in general a complex
number. When $q=1$, the boson commutation relations of the
harmonic oscillator are recovered.

A possible Hamiltonian for the system (\ref{qanh}) is:
\begin{equation}
H=\frac{1}{2}(aa^{\dagger }+a^{\dagger }a)=
\frac{1}{2}\mathcal{C}+\frac{1}{2} \frac{q^{\hat{n}
+1}+q^{\hat{n}}-2}{q-1} .\label{hamq}
\end{equation}
where  the Casimir operator $\mathcal{C}$ can be written in the
form:
\begin{equation}
\mathcal{C}=aa^{\dagger }+a^{\dagger
}a-\frac{q^{{\hat{n}}+1}+q^{\hat{n}}-2}{q-1} .
\end{equation}
and  satisfies the commutation relations,
\begin{equation}
\lbrack \mathcal{C},a]=[\mathcal{C},a^{\dagger
}]=[\mathcal{C},\hat{n}]=0
\end{equation}

As shown in \cite{angelova:JPA}, the anharmonic bosons (\ref{anh})
can be obtained from the $q$-bosons (\ref{qanh})  for real values
of the deformation $q$ close to 1, $q<1$, using  an expansion of
first order of $q$ in terms of a parameter $p$, $p\equiv 1/(1-q)$,
\begin{equation}
q^{\hat{n}}=1-\frac{\hat{n}}{p}~~.  \label{qqanh}
\end{equation}
where $1/p{\ll}1$. If we now substitute the approximation for
$q^{\hat{n}}$ from equation (\ref {qqanh}) in the commutation
relations (\ref{qanh}) and identify the parameter $p$ with
$\mathcal{N}/2$, $\hat{n}$ with $\hat{v}$ and the creation and
annihilation operators $a,a^{\dagger }$, with $b,b^{\dagger }$
respectively, we recover the $su(2)$ anharmonic relations
(\ref{anh}).

For $q\leq1$, the case of harmonic and anharmonic vibrations in
molecules and solids is retrieved. The form (\ref{anh}) of the
$su(2)$ commutation relations can be considered as a deformation
of the usual (harmonic oscillator) commutation relations, with a
quantum deformation parameter $p=\mathcal{N}/2$.

This gives  a physical interpretation of the quantum deformation.
The quantum deformation parameter $p$ is the fixed number
$\mathcal{N}_{0}$ of the anharmonic bosons in the oscillator.
Using the relation between the fixed number of anharmonic bosons
$\mathcal{N}$ and the characteristics of the Morse potential
(\ref{ND}), we can conclude  that the quantum deformation
parameter is also determined by the depth, the width and in
general the shape of the Morse potential well. For the molecule
$^{1}$H$^{35} $Cl, $p=28$ and   $q=27/28$.
Substituting $\mathcal{N}_{0}=p$ in the expressions (%
\ref{pf}), (\ref{U}),  (\ref{C})and (\ref{v}), we obtain  the
thermodynamic properties of diatomic molecules  in terms of the
deformation parameter $p$.

The case $q>1$ is also very interesting and is related to
Bose-Einstein condensation and superfluidity
\cite{monteiro:physrevlet,ubri,jellal}.

\subsection{q-Bosons at roots of unity}

In  \cite{angelova:JPA}, we have presented a new algebraic
realization of the $q$-bosons , for the case of $q$ being a root
of unity, $q^N\equiv1$ (integer $N>1$), which corresponds to a
periodic structure described by a finite-dimensional
representation. Note that the integer  $N$ in this section is not
related in general to the total number of bosons $\mathcal{N}$ in
the previous sections.

We consider the  operator $K\equiv q^{\hat{n}}$  with commutation
relations,
\begin{equation}
\lbrack a,a^{\dagger }]=K\ ,\quad K\,a=q^{-1}\,a\,K\ ,\quad
K\,a^{\dagger }=q\,a^{\dagger }\,K.
\end{equation}

$|\rangle \,$ is  the vacuum state which is annihilated by the
operators lowering the boson number and is an eigenvector of the
number operator:
\begin{equation}
a\,|0\rangle \,=0\ ,\qquad K\,|\rangle \,=\, q^\mu |0\rangle\, ,
\qquad ({\rm or}\ \ \hat{n}|\rangle \,=\mu |0\rangle) \,
\end{equation}
where $\mu $ in the generic case   is an arbitrary complex number.
The states of the system are built by applying the operators
raising the boson number:
\begin{equation}
|k\rangle \ \equiv \ (a^{\dagger })^{k}\,|0\rangle \,.
\end{equation}
The action of the algebra on the basis $|k\rangle \,$ is:
\begin{eqnarray}
&&K|k\rangle \,=q^{\mu +k}|k\rangle \, \label{action} \nonumber
\\ &&a\,|k\rangle \,=q^{\mu }\,\frac{q^{k}-1}{q-1}\,|k-1\rangle
\, \nonumber \\ &&a^{\dagger }\,|k\rangle \,=|k+1\rangle .
\end{eqnarray}
The representation space,  $V_{\mu }$, is infinite-dimensional for
a  generic deformation parameter.

Now, let $ q$ is a nontrivial root of unity, {\textit i.e.},
$q^{N}=1$ for integer $N>1$. In this case we have:
\begin{equation}
a\,|N\rangle \,=q^{\mu }\,\frac{q^{N}-1}{q-1}\,|N-1\rangle \,=0.
\label{trunk}
\end{equation}
All states $\ |k\rangle \,\ $ with $k\geq N$ form an infinite
dimensional invariant subspace, $I_{\mu}$. We then obtain a
finite-dimensional representation space, which is the
factor-space, $ F_{\mu}=V_{\mu}/ I_{\mu}$, with dimension $N$.
Considering the action of the algebra on the states, $|k+mN\rangle
$, for fixed $k<N$ and for all non-negative integer $m$, we have
shown in \cite{angelova:JPA} that the structure is periodic,
{\textit i.e.}, the action of the algebra on all states
$|k+mN\rangle $ (for fixed $k$) coincides. Thus, it is sufficient
to consider the states $|k\rangle $ with $k<N$. Let us denote
these identified states by $\widetilde{|k\rangle }$, $k=0,\dots
,N-1$. They form a finite-dimensional representation space $\tilde
F_{\mu }$ with dimension $N$. The action of the algebra on these
states is:
\begin{eqnarray}
&&K\,\widetilde{|k\rangle }=q^{\mu }\widetilde{|k\rangle }
\label{faction} \nonumber \\ &&a\,\widetilde{|k\rangle }=q^{\mu
}\frac{q^{k}-1}{q-1}\,\widetilde{ |k-1\rangle } \nonumber \\
&&a^{\dagger }\,\widetilde{|k\rangle }=\widetilde{|k+1\rangle }\
,\qquad k<N-1 \nonumber \\ &&a^{\dagger }\,\widetilde{|N-1\rangle
}=\widetilde{|0\rangle }, \ \ \widetilde{|0\rangle
}\equiv\widetilde{ |N\rangle }. \label{alg}
\end{eqnarray}

This represents a finite-dimensional system, where  the boson
number lowering operator acts in the usual way, (in particular, it
annihilates the vacuum state $\widetilde{|0\rangle }$), but the
boson raising operator acts cyclicly. It  has a non-zero action on
all states and  the vacuum state may be obtained not only by the
action of the lowering operator but also by the action of the
raising operator producing a jump from $\widetilde{|N-1\rangle }$
to $\widetilde{|0\rangle }$. One realization of this operator is a
two-level system, obtained for $N=2$ (equivalent to $q=-1$,
fermions). For $N>2$, systems with finite number of levels and
population inversion are illustrations of possible action of these
operators. In \cite{DHS} the objects  "quons" ($q^N=1$) are
introduced as  interpolating between fermions ($N=2$) and
bosons($N\rightarrow\infty$).

For large $N$ periodicity of the type decribed above appears in
crystals. We have shown  that it represents the periodic boundary
conditions, first proposed by Born and von K\'{a}rm\'{a}n
\cite{born:physz}. The periodic boundary conditions
\cite{corn:group1} are imposed on the translational symmetry,
which strictly speaking is a property of an infinite crystalline
lattice, to allow its use for finite crystals. The periodic
boundary conditions determine the number of the allowed
wave-vector states in the Brillouin zone model and imply
additional selection rules on certain frequencies.

For  the classic example of a linear lattice of identical
particles with periodic boundary conditions, the equilibrium
positions of the particles are given by
$\mathbf{t}_{n}=n\mathbf{t}\mathbf{,\;\;}n=0,1\mathbf{,\ldots
,}N-1$, and the periodic boundary condition requires
$\mathbf{t}_{N}\equiv \mathbf{t}_{0}\equiv \mathbf{0.}$,  where
$\mathbf{t}$ is the vector of primitive translations and $N$ is a
large positive number.

The symmetry operations of the linear lattice form a cyclic finite
group of\ order $N $ with a generator, the primitive translation
$\{E|\mathbf{t}\}$. Here, the Seitz notation is used to represent
a translation and \ $E$ is the identity, $E\equiv \{E|0\}$.

The $n$-th element of the group is
\begin{equation}
\{E|t\}^{n}=\{E|\mathbf{t}_{n}\},\;n=1,2,\ldots ,N-1.
\end{equation}
The product of two elements of the group is an element of the
group
\begin{equation}
\{E|t_{m}\}\{E|t_{n}\}=\{E|t_{m+n}\},\;m,n=1,2,\ldots ,N-1, \;
\end{equation}
where $m+n\equiv (m+n)\, ({\rm mod}\,N)$. The identity is
\begin{equation}
\{E|\mathbf{t}\}^{N}\equiv \mathbf{\{}E|\mathbf{t}_{0}\}\equiv \{E|%
\mathbf{0}\}.
\end{equation}
Using the action (\ref{alg}) of the operator $a^{\dagger }$ on the
states $\widetilde{|k\rangle }$, we have shown that the raising
operator $a^{\dagger }$ is isomorphic to the generator
$\{E|\mathbf{t}\}$.

Thus, the symmetry group of the lattice with periodic boundary
conditions is isomorphic to the finite cyclic group of order $N$
with a generator, the operator $a^{\dagger }$. This group can be
used with the other symmetry operations of one-dimensional
crystalline or polymer Hamiltonians. The boundary conditions can
be generalised for the three-dimensional case by introducing
raising operators for each dimension.

\section{Conclusion}

In this paper, we have used the  classical and quantum algebras to
study the vibrational thermodynamic properties of molecules and
solids. The algebraic approach is applied to those thermodynamic
properties of diatomic molecules which at high temperatures
strongly depend on the anharmonic effects. The vibrational
thermodynamic functions, such as partition function, mean energy
and specific heat, are derived in terms of the parameters of the
algebraic model and their properties are discussed.

The application of a $q$-algebra to physical problems is often
lacking an appropriate interpretation for the deformation
parameters and often applications are carried out where
generalisation to $q$-deformed versions of well known models are
made with no simple interpretation. We have shown that the $1/p$
approximation leads to the $su(2)$ algebra and to an
interpretation of $p$ in terms of the Morse potential
anharmonicity. We have found a physical interpretation of the
quantum deformation, showing that the deformation parameter is
related to the fixed number of anharmonic bosons and the shape of
the Morse potential. We have found  that all  vibrational
thermodynamic properties (\textit{e.g.} mean energy, specific
heat, mean number of anharmonic bosons) of the molecules depend on
the corresponding quantum deformation parameter.

A new application of the $HW_q$ algebra when $q$ is a root of
unity is discussed, which gives a periodic structure described by
a finite-dimensional representation. The raising operator
belonging to this structure generates a group isomorphic to the
symmetry group of a linear lattice with periodic boundary
conditions. The latter may provide a useful framework for the
deformation of crystalline or polymer Hamiltonians.

\section*{Acknowledgements}
The results in this work are obtained  in collaboration with A.
Frank and V. Dobrev. The author would like to thank  the
Organising Committee of the IXth International Conference on
Symmetry Methods in Physics for inviting her to give a plenary
talk.

\end{document}